\documentclass[aps,pra,twocolumn,amsmath,amssymb,bibnotes]{revtex4}

\usepackage{color,epsfig,rotating,graphicx,subfigure}

\usepackage[english]{babel}

\usepackage{dsfont}
\usepackage{textcomp}
\newcommand{\kb}{k_{\rm B}}

\begin{document}
\title{On the heat transfer across a vacuum gap mediated by Casimir force}

\date{\today}

\author{Svend-Age Biehs$^{1,*}$, Achim Kittel$^1$, and Philippe Ben-Abdallah$^2$}
\affiliation{$^1$ Institut f{\"u}r Physik, Carl von Ossietzky Universit{\"a}t, D-26111 Oldenburg, Germany}
\affiliation{$^2$ Laboratoire Charles Fabry, UMR 8501, Institut d’Optique, CNRS, Université Paris-Saclay, 2 Avenue Augustin Fresnel, 91127 Palaiseau Cedex, France}
\email{ s.age.biehs@uni-oldenburg.de} 

%%%%%%%%%%%%%%%%%%% abstract and OCIS codes %%%%%%%%%%%%%%%%

\begin{abstract}
Following a recent report on experiments it has been claimed that a phonon heat transfer through a vacuum gap between two  solids due to quantum fluctuations had been measured. Here we make a theoretical analyzis of this mechanism and demonstrate that the Casimir force driven heat flux is at least 15 orders of magnitude smaller than the near-field heat flux between two solids at any given separating distance. Moreover, we provide compelling arguments that the measures reported are no more and no less a measure of the virtual temperatures of vibrating modes which does not allow an access to the Casimir force driven heat flux. 

\end{abstract}

\maketitle

% 
% Motivation/Einleitung des Effekts und des Experiments
%

%\section{Introduction}

Recently, a report on experiments have been published~\cite{FongEtAl} which have been interpreted to demonstrate ``heat transfer driven by quantum fluctuations using nanomechanical devices''. This startling phenomenon, in its turn, allegedly ``has practical implications to thermal management in nanometre-scale technologies'' and ``paves the way for the exploitation of quantum vacuum in energy transport at the nanoscale''~\cite{FongEtAl}. Or as summarized in a subsequent comment~\cite{PhysicsToday}: ``Phonon heat transfer through vacuum could have implications for managing heat in integrated circuits. The mechanism may provide a new way to intentionally dissipate heat in high-density transistor circuits, and it is an important consideration for keeping close elements thermally isolated in, for example, optical communication devices, which are sensitive to temperature cross talk. On a fundamental level, the conventional three methods of heat transfer-conduction, convection, and radiation --- must now become four.''

In fact, in Ref.~\cite{FongEtAl} it is claimed that a maximum heat flux of $P_{\rm Casimir} = 6.5\times10^{-21}\,{\rm W}$ between two gold coated Si$_3$N$_4$ membranes (see Fig.~2(a) in \cite{FongEtAl}) could be measured which is due to the coupling of both membranes by the Casimir force. This ``phonon heat transfer through quantum fluctuations'' is claimed to result in an apparent cooling (heating) of the hot (cold) membrane from $312.5\,{\rm K}$ ($287\,{\rm K}$) at a separation distance of $800\,{\rm nm}$ to the final temperature of about $305\,{\rm K}$ at a separation distance of $350\,{\rm nm}$ (see Fig.~3(a) in \cite{FongEtAl}). As discussed by the authors the heat transfer by thermal radiation would lead to a temperature change of only 0.02K even when the nowadays well studied near-field enhancement~\cite{Fiorino} is included. Seemingly, the Casimir effect driven heat transfer is much more important than the heat transfer due to near-field thermal radiation which is estimated by the authors in Ref.~\cite{FongEtAl} to have a maximal value of $1.4\,{\rm W}/{\rm m}^2 {\rm K}$ at $350$nm distance.

However, multiplying this value by the surface area of the membranes of $330\times330\,\mu {\rm m}^2$ and the temperature
difference of about $25\,{\rm K}$ one obtains a near field radiative heat flux of $P_{\rm rad} = 3.5\, \mu {\rm W}$ which is 15 orders of magnitude larger than the heat flux $P_{\rm Casimir}$ due to the Casimir force. Therefore, one would expect that the Casimir force driven heat transfer would lead to a temperature change within the membranes of $0.02\times10^{-15} \,{\rm K}$
and hence is negligibly small compared to the radiative heat flux. But this seems to be in strong contradiction to the
experimental results allegedly suggesting that the Casimir force driven heat flux leads to a temperature change of several
Kelvin. 

In the present Letter, we disentangle this apparent contradiction by providing a detailed theoretical analysis of this effect. We
demonstrate that, on one hand, Casimir force driven heat flux is orders of magnitude smaller than the heat flux exchanged by photons 
between two near-field coupled solids and, on the other hand, the measured quantity reported in~\cite{FongEtAl} is not an actual temperature 
change of the membranes due to a Casimir force driven heat flux but a measurement of the change in mean squared displacement of the membranes interpreted as a mode temperature by the authors. These mode temperatures resemble the temperature of the heat bath only if the membranes are tightly coupled to the heat bath and if they are decoupled from any surrounding, i.e. in thermal equilibrium with the heat bath. Finally, we discuss the question whether the experiment in Ref.~\cite{FongEtAl} indeed proves the presence of a phonon heat transfer through quantum fluctuations.

%Recently, experimental results have been reported~\cite{FongEtAl} on a heat transfer mechanism mediated by thermal vibrations through a vacuum gap between two  solids due to quantum fluctuations, paving so a new avenue for thermal management at nanoscale~\cite{PhysicsToday}. True thermal analog of the famous Casimir force this fourth channel for heat transfer is being considered as a potential solution to dissipate heat at nanoscale in high-density electric circuits and to  keep close elements thermally isolated, for example, optical communication devices, which are sensitive to temperature cross talk. 
%In the present Letters we make a theoretical anayzis of this effect and demonstrate that on one hand it is orders of magnitude smaller than the heat flux exchanged by photons tunneling between two solids whatever their separation distance is and on the other hand  the measured quantity reported in~\cite{FongEtAl}  is not a  temperature changes of solids due to a Casimir heat flux but neither more nor less than a measure of the virtual temperature of their vibrating modes that is a meaure of mean squared  displacements. 

%\section{Coupled harmonic oscillators}

\begin{figure}
  \includegraphics[width=0.35\textwidth]{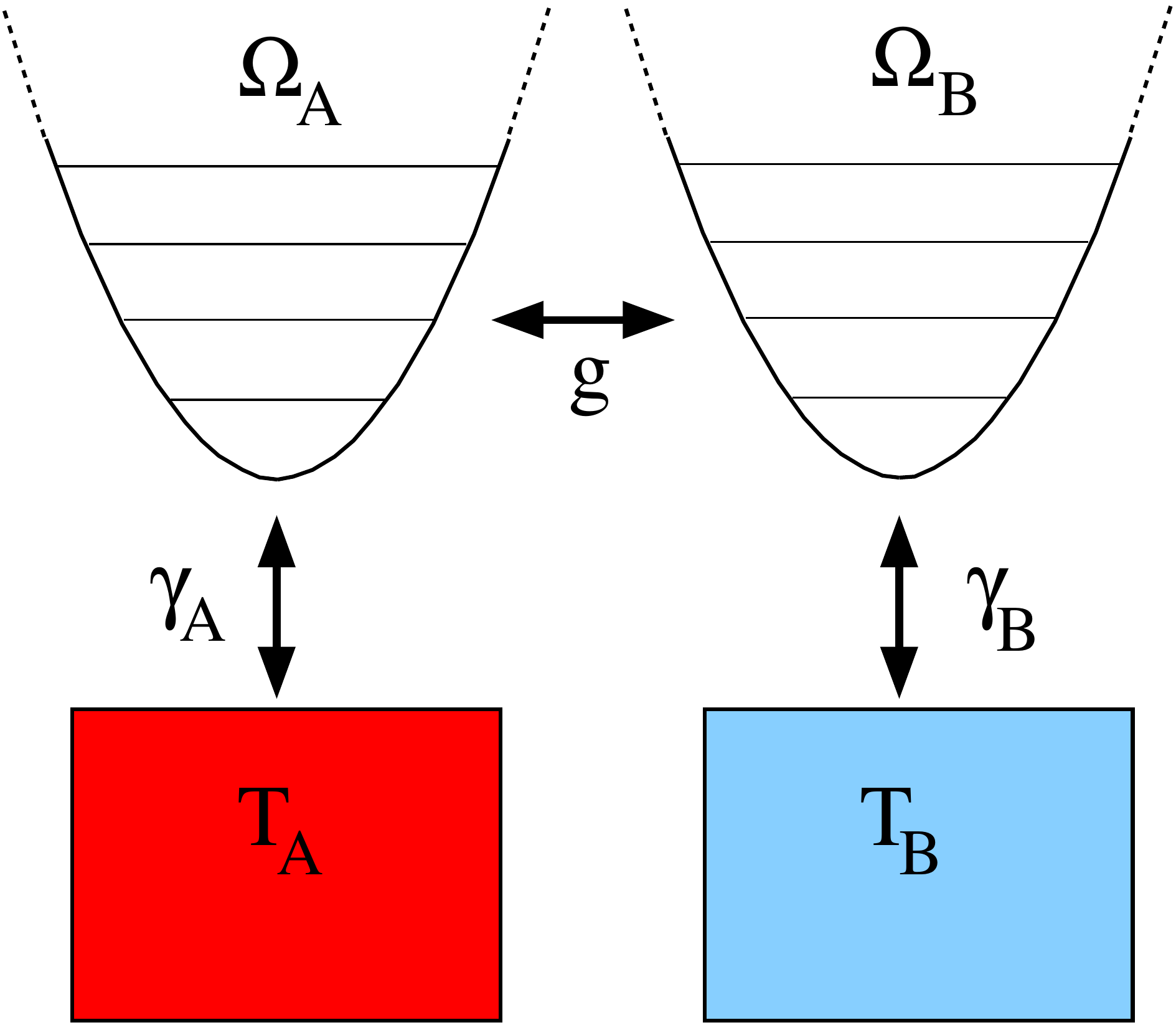}
  \caption{Sketch of the two harmonic oscillators with frequencies $\Omega_A = \Omega_B \equiv \Omega$ coupled to two heat baths at temperatures $T_A$ and $T_B$ with coupling strengths $\gamma_A$ and $\gamma_B$. The two oscillators are coupled by a linearized force described by the coupling strength $g$ which is here the Casimir force between the membranes.}
  \label{Fig:Oscillators}
\end{figure}

To start let us consider a simple system described by two harmonic oscillators $A$ and $B$ as depicted in Fig.~\ref{Fig:Oscillators}. The oscillation frequencies of both oscillators are equal as in the experiment, i.e.\ we set $\Omega_A = \Omega_B \equiv \Omega$. The phenomenological damping rates $\gamma_A = \Omega/ 2 Q_A$ and $\gamma_B = \Omega/ 2 Q_B$ are determined by 
the quality factors of the membranes which are in the experiment $Q_A = 4.5\times10^4$ and  $Q_A = 2\times10^4$. They describe the coupling strength of the vibrational oscillator modes to the heat baths provided by the atoms in the membrane and membrane frame which are held at different temperatures $T_A$ and $T_B$. We further assume that the two membranes are coupled via a force whose strength is described by a coupling constant $g$.

 In a full quantum description~\cite{SAB} the mean occupation number $\langle \hat{a}^\dagger \hat{a} \rangle$ (resp. $\langle \hat{b}^\dagger \hat{b} \rangle$) of two coupled  harmonic oscillators in thermal equilibrium  with two thermal baths at temperatures $T_A$ and $T_B$ reads under the standard Born-Markov assumption 
%\begin{equation}
%  \langle \hat{a}^\dagger \hat{a} \rangle = \frac{g^2 \frac{\gamma_A n_A + \gamma_B n_B}{\gamma_A + \gamma_B} + \gamma_A \gamma_B n_A}{g^2 + \gamma_A \gamma_B}. \label{Eq:adaggera}
%\end{equation}
\begin{align}
  \langle \hat{a}^\dagger \hat{a} \rangle &= \frac{g^2 \frac{\gamma_A n_A + \gamma_B n_B}{\gamma_A + \gamma_B} + \gamma_A \gamma_B n_A}{g^2 + \gamma_A \gamma_B}, \label{Eq:adaggera}\\
  \langle \hat{b}^\dagger \hat{b} \rangle &= \frac{g^2 \frac{\gamma_A n_A + \gamma_B n_B}{\gamma_A + \gamma_B} + \gamma_A \gamma_B n_B}{g^2 + \gamma_A \gamma_B}
\end{align}
where $\hat{a}$  (resp. $\hat{b}$) denotes the ladder operator of oscillator $A$  (resp. $B$) which fulfill the standard commutation relations $[\hat{a},\hat{a}^\dagger] = 1$ (resp. $[\hat{b},\hat{b}^\dagger] = 1$) and 
\begin{equation} 
  n_{A/B} = \frac{1}{\exp(\hbar \Omega /\kb T_{A/B}) - 1}
\end{equation}	
is the Bose distribution function. Note, that in the experiment the frequency of the oscillations of the membranes are $\Omega/2 \pi = 191.6\,{\rm kHz}$ so that $\hbar \Omega \ll \kb T_{A/B}$ and therefore the classical limit $n_{A/B} \approx \frac{\kb T_{A/B}}{\hbar \Omega}$ is applicable and will be used further.

Before we discuss the heat flow between the coupled harmonic oscillators let us consider two very important limiting cases. In the weak coupling limit ($g \ll \gamma_A, \gamma_B$) and the low frequency regime ($\hbar \Omega \ll \kb T_{A/B}$) it holds
\begin{equation}
   \langle \hat{a}^\dagger \hat{a} \rangle = n_A \approx \frac{\kb T_{A}}{\hbar \Omega}\quad \text{and} \quad
   \langle \hat{b}^\dagger \hat{b} \rangle  = n_B\approx \frac{\kb T_{B}}{\hbar \Omega}. \label{Eq:WCL}
\end{equation}
Hence, in this limit the oscillators are thermalized by their heat baths at the corresponding bath temperature. Now, in the strong coupling limit ($g \gg \gamma_A, \gamma_B$) we have
\begin{equation}
  \langle \hat{a}^\dagger \hat{a} \rangle \approx \langle \hat{b}^\dagger \hat{b} \rangle\approx  \frac{\kb}{\hbar \Omega} \frac{\gamma_A T_A + \gamma_B T_B}{\gamma_A + \gamma_B},
\label{Eq:SCL}
\end{equation}
Thus, in the strong coupling limit the mean occupation numbers are the same for both oscillators.
It must be noted that the two oscillators are not in equilibrium at the same temperature, but their mean occupation number is in some sense the weighted mean value of the occupation numbers of the decoupled oscillators.

%\section{Power flow between the harmonic oscillators}

The heat flowing between the two reservoirs which are coupled via the oscillators by the coupling constant $g$ in steady state reads in the classical limit~\cite{SAB}
\begin{equation}
  P_{A \rightarrow B} \approx 2 g^2 \frac{\gamma_A \gamma_B}{(\gamma_A + \gamma_B)(g^2 + \gamma_A \gamma_B)} \kb (T_A - T_B). 
\label{Eq:PAB}
\end{equation}
From this expression, it can be seen that any coupling constant $g$ leads to a heat transfer between the reservoirs  as predicted by several authors in recent theoretical studies~\cite{Pendry,Joulain,Budaev} considering different models for phonon tunneling including also Casimir force mediated heat fluxes. In the strong coupling limit the power simplifies into the following form
\begin{equation}
   P_{A \rightarrow B} \approx 2 \frac{\gamma_A \gamma_B}{\gamma_A + \gamma_B} \kb (T_A - T_B).
\end{equation}
Therefore, the heat flowing between the reservoirs is analog to the heat flowing through two thermal resistances $R_A = (\kb \gamma_A)^{-1}$ and $R_B = (\kb \gamma_B)^{-1}$ in series.  From this simple model and using the values from the experiment in~\cite{FongEtAl}, we get a power of $P \approx 6.5\times10^{-21}\,{\rm W}$ which precisely coincides with the maximum value  $P_{\rm Casimir}$ deduced from the results of the experiment in Ref.~\cite{FongEtAl}. This value will be achieved no matter however the strong coupling limit is imposed and it is extremely small. For example, the heat conductance of single molecules is on the order of $18 \times 10^{-12}\,{\rm W}/{\rm K}$~\cite{Reddy}. This means, for a temperature difference of $25\,K$ applied at both ends of the molecule one would have a heat flowing through the molecule at a rate of about $4.5\times 10^{-10}\,{\rm W}$, i.e.\ a value which is 11 orders of magnitude larger than $P_{\rm Casimir}$.

%\section{Definition of mode temperatures}

Before going further in the analysis of this Casimir force driven heat flux it is now important to recall some standard features of harmonic oscillators. Here we focus, for clarity reasons, only on oscillator $A$, but of course corresponding relations also hold for the oscillator $B$. Quantum mechanics predicts that its  mean kinetic and mean potential energy reads~\cite{SAB} 
\begin{align}
  \langle E^{\rm kin}_A \rangle &= \frac{1}{2} m_A \langle \dot{\hat{x}}_A^2 \rangle =  \langle \hat{a}^\dagger \hat{a} \rangle  \frac{\hbar \Omega}{2}, \label{Eq:Ekin} \\
  \langle  E^{\rm pot}_A \rangle  &= \frac{1}{2} m_A \Omega^2 \langle \hat{x}_A^2 \rangle = \langle \hat{a}^\dagger \hat{a} \rangle  \frac{\hbar \Omega}{2}.
\end{align}
In the weak coupling limit  Eq.~(\ref{Eq:WCL}) allows us to obtain the simplified expressions
\begin{equation}
  \langle E^{\rm kin}_A \rangle  = \langle E^{\rm pot}_A \rangle \approx  n_A  \frac{\hbar \Omega}{2} \approx \frac{\kb T_A}{2} 
\end{equation}
showing the classical result where each degree of freedom contributes up to $\kb T_A / 2$ to the mean energy in equilibrium.
Of course, in this limit the oscillator is in equilibrium with its heat bath completely and decoupled from its surrounding. On the other hand,
in the strong coupling limit we have with Eq.~(\ref{Eq:SCL}) 
\begin{equation}
  \langle E^{\rm kin}_A \rangle  = \langle E^{\rm pot}_A \rangle 
                                 \approx \frac{\gamma_A}{\gamma_A + \gamma_B} \frac{\kb T_A}{2} + \frac{\gamma_B}{\gamma_A + \gamma_B} \frac{\kb T_B}{2}.
\end{equation}
As already noted before, in this coupling regime the oscillators are not in equilibrium with their baths at all/anymore and each bath contributes up to a weighed value of $\kb T_A / 2$ plus $\kb T_B/2$  by the relative coupling of the oscillators to the baths. 

Despite the fact that due to the coupling the oscillators are in general not in thermal equilibrium, in Ref.~\cite{SAB} the mode temperature $T_A'$ is defined by the authors as to be
\begin{equation}
  \langle E^{\rm kin}_A \rangle  = \langle E^{\rm pot}_A \rangle \approx  n_A  \frac{\hbar \Omega}{2} \approx \frac{\kb T_A'}{2}, 
\label{Eq:DefModeTemp}
\end{equation}
for all values of $g$. Setting this definition equal to Eq.~(\ref{Eq:Ekin}) we have
\begin{equation}
  T_A' = \frac{\hbar \Omega}{\kb} \langle \hat{a}^\dagger \hat{a} \rangle
\end{equation}
which provides in the classical limit of Eq.~(\ref{Eq:adaggera}) a relation to the bath temperatures which can be expressed as
\begin{equation}
  T_A' = T_A + \frac{g^2 \gamma_B (T_B - T_A)}{(\gamma_A + \gamma_B)(g^2 + \gamma_A \gamma_B)}.
\end{equation}
Hence, in the weak coupling limit ($g \ll \gamma_A, \gamma_B$) one has $T_A' = T_A$ and in the strong coupling limit ($g \gg \gamma_A, \gamma_B$) one has obviously
\begin{equation}
  T_A' \approx T_B' \approx \frac{\gamma_A}{\gamma_A + \gamma_B} T_A + \frac{\gamma_B}{\gamma_A + \gamma_B} T_B.
\label{Eq:ModeBathT}
\end{equation}
Therefore, the mode temperatures associated with the mean square displacement only coincide with the bath temperatures in the weak coupling limit. In that case the mode temperatures are resembling real temperatures which could also be measured with a thermometer. In contrast, in the strong coupling limit the values of the mean square displacements of both oscillators become equal. This just reflects the fact that in the strong coupling regime the mean occupation numbers of the two oscillators become equal due to the coupling. 
If we insert the values of the experiment in Ref.~\cite{FongEtAl} we obtain $T_A' \approx T_B' \approx 304.7\,{\rm K}$ which agrees very well with the value measured at $350\,{\rm nm}$ separation distance in~\cite{FongEtAl}. Note, that although the mode temperatures, i.e. the value of the mean square displacement, seem to become equal in the strong coupling limit, the bath temperatures do not change at all. Due to the extremely small value of $P\approx 6.5\times10^{-21}\,{\rm W}$ it is, in principle, clear that equilibrating the two heat baths is practically impossible. But from the theoretical point of view, it is right from the start excluded that the bath temperatures change, which is obviously also the assumption underlying the data fitting procedure in Ref.~\cite{FongEtAl}.

% 
% Ist das Experiment tatsächlich eine Wärmestrommessung 
%

%\section{Heat flux measurement}
But how could then a ``heat flux'' been measured? 
To answer this question let us compare Eq.~(\ref{Eq:ModeBathT}) and Eq.~(\ref{Eq:PAB}). By doing so we find
\begin{equation}
  P_{B \rightarrow A} = 2 \gamma_A \kb (T_A - T_A')
                      = - 2 \gamma_B \kb (T_B - T_B')
\label{Eq:PBAprime}
\end{equation}
This means that the heat flux exchanged between the heat baths through the coupled oscillators is the same as the heat flux
between the bath and the oscillator which is assumed to be equilibrated at the mode temperature $T_A'$ or $T_B'$. Here, the 
thermal resistance is $(\kb \gamma_A)^{-1}$ or $(\kb \gamma_B)^{-1}$ respectively. Using the values from the experiment we 
find therefore for the conductances  $\kb \gamma_A = 1.9\times10^{-22}\,{\rm W/K}$ and $\kb \gamma_B = 4.2\times10^{-22}\,{\rm W/K}$ which are very small. This is inevitable in the experiment, because in order achieve the strong coupling regime ($g \gg \gamma_A, \gamma_B$) it is necessary that the conductance $\kb g$ due to the Casimir force between the oscillators becomes much larger than $\kb \gamma_A$ and $\kb \gamma_B$. 

From Eq.~(\ref{Eq:PBAprime}) it becomes clear that to measure a heat flux between the reservoirs it suffices to determine either
the mode temperature $T_A'$ or $T_B'$ together with the bath temperatures $T_A$ or $T_B$. Actually, the only thing which needs
to be measured is $T_A'$ or $T_B'$, because $T_A$ and $T_B$ are fixed and they do not change during the experiment. But it
is not necessary to use the concept of mode temperatures, because what really counts is simply  $\langle \hat{a}^\dagger \hat{a} \rangle$ or equivalently $\langle E^{\rm pot}_A \rangle$ or $\langle \hat{x}_A^2 \rangle$. And in the experiment in Ref.~\cite{FongEtAl} exactly the quantities $\langle \hat{x}_A^2 \rangle$ and $\langle \hat{x}_B^2 \rangle$ have been measured. This is done by measuring the values for the quadrature of the displacements of the membranes which allow to determine the distribution of the displacement. This distribution can be mapped to an energy distribution which, under the assumption it is excited by a heat bath, can be fitted by a Boltzmann distribution $\propto \exp(- E^{\rm total}_{A/B} /\kb T'_{A/B})$ allowing to determine the mode temperature. This gives the impression that the membranes are equilibrated at their temperatures $T_A'$ and $T_B'$ so that the displacement measurement corresponds to a temperature measurement. But it must be kept in mind that only in the weak coupling limit ($g \ll \gamma_A, \gamma_B$) the vibrational modes are equilibrated to their bath temperatures and in general the mode temperatures have no thermodynamic meaning. Actually the displacement of the membranes or oscillators has the Gaussian property so that in phase space the quadrature can be fitted by a two dimensional Gaussian distribution. The distribution of the displacement itself is a one dimensional Gaussian $\propto  \exp(- x^2_{A/B}/2 \langle \hat{x}^2_{A/B}\rangle)$ which resembles the Boltzmann distribution. A fit with this function yields $\langle \hat{x}^2_{A/B}\rangle$, i.e.\ the quantity which is really measured in the experiment. With relation (\ref{Eq:DefModeTemp}) we canassociate mode temperatures $T_A'$ and $T_B'$, but this step is not necessary at all but even misleading for the interpretation. 

%\section{Conclusion}

The question whether the measurement in Ref.~\cite{FongEtAl} is a heat flux measurement might sound somewhat philosophical. However, 
it is of uttermost importance for physics in general to ask such questions to draw the right conclusions and discuss possible consequences. One of the main tasks of physics
is to verify or falsify a theoretical model by an experiment or to measure effects which needs to be explained by a theory.
Here, we need to give an answer to the question: Does the experiment prove that there is a heat flux which can be described
by the coupled oscillator model? If we now reflect the experiment, we might come to the conclusion that 
any kind of measurement of the Casimir force for two membranes with fixed temperatures $T_A \neq T_B$ would give the coupling 
constant $g$ which is completely sufficient to calculate $P_{B \rightarrow A}$ as can be seen from Eq.~(\ref{Eq:PAB}) assuming of course
that $\gamma_A, \gamma_B$, $T_A$, and $T_B$ are known. In the experiment $\langle \hat{x}_A^2 \rangle$
has been measured which is of course a mean to determine $g$. Therefore, one might argue that the measurement 
in Ref.~\cite{FongEtAl} is simply a Casimir force measurement for two membranes having different bath temperatures. The heat flux
 itself is not measured by any means. There is no measurement of a temperature change of the membranes only a measurement of a virtual mode temperature which is nothing else than $\langle \hat{x}_A^2 \rangle$ and which cannot be measured by an actual thermometer. Taking the value $P_{\rm Casimir} = 6.5\times10^{-21}\,{\rm W}$ calculated for the largest possible Casimir force driven heat flux in the experimental setup it even seems to be impossible to really measure such a heat flux because the temperature change of the membranes would be vanishingly small compared to the temperature change due to the radiative heat flux. Hence, the experimental setup cannot measure any heat flux, but it concludes from the measured Casimir force effect on $\langle \hat{x}_A^2 \rangle$ that there must be a heat flux. This is a rather indirect ``proof'' which completely hinges on the validity of the theoretical model for $P_{B \rightarrow A}$ which is a priori assumed to be valid. Or differently stated, without the equation for $P_{B \rightarrow A}$ the experiment could not give any number for the Casimir force driven heat flux and it is hence no independent measurement that Casimir force driven heat flux or a proof of the validity of the model for $P_{B \rightarrow A}$.

\begin{figure}
  \includegraphics[width=0.45\textwidth]{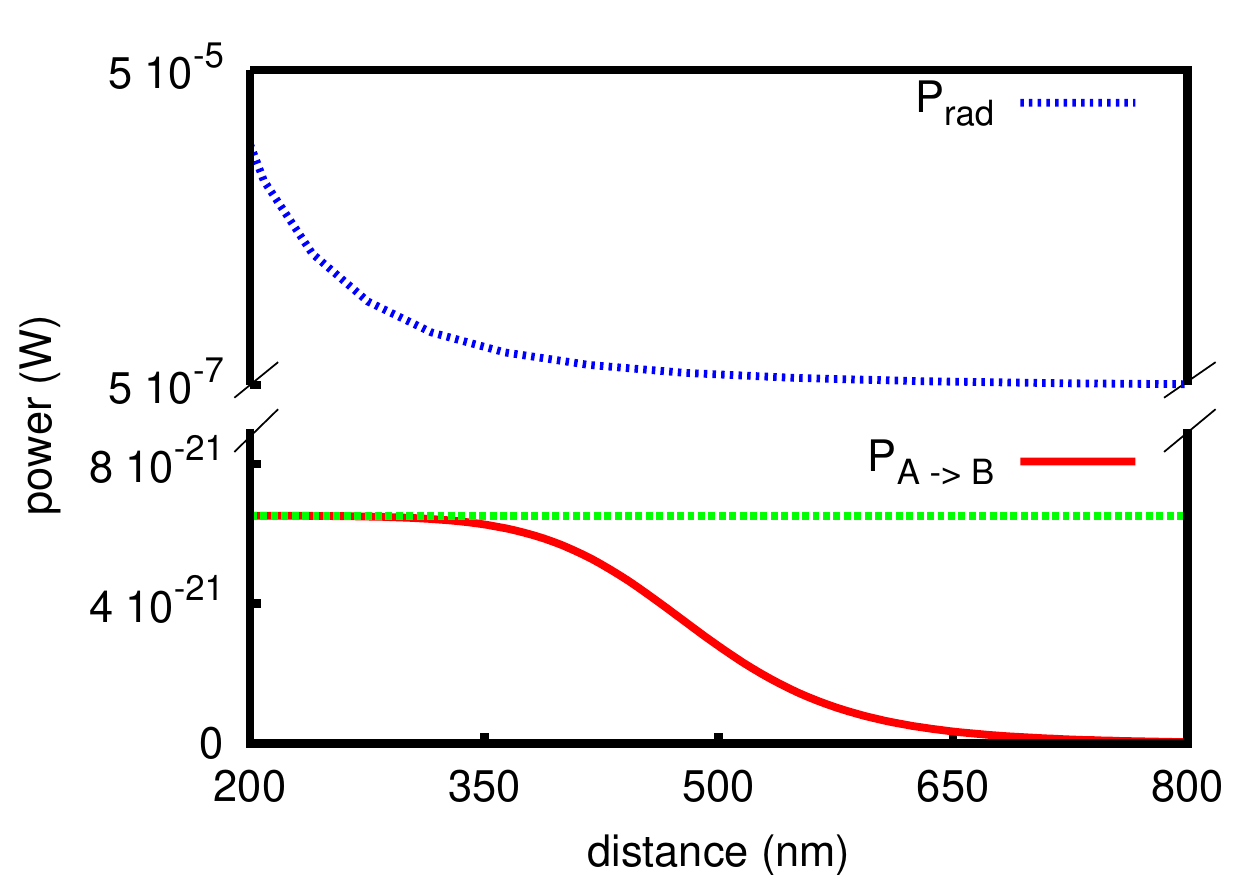}
  \caption{Comparison of heat flux $P_{A \rightarrow B}$ mediated by the Casimir force using Eq.~(\ref{Eq:PAB}) with the  values from the experiment for $T_{A/B}$, $\gamma_{A/B}$ and the measured coupling constant $g$ from\cite{FongEtAl} with the radiative heat flux $P_{\rm rad}$ between the membranes using fluctuational electrodynamics~\cite{PvH}. It can be obviously seen that for small distances the strong coupling regime for $P_{A \rightarrow B}$ is reached and the flux converges to $ 6.5\times10^{-21}$ (horizontal line). Note, that the radiative heat flux is by 14 to 16 orders of magnitude larger than the Casimir force mediated heat flux.}
  \label{Fig:HeatFlux}
\end{figure}

To summarize, we have demonstrated that the experimental results reported in Ref.~\cite{FongEtAl} do not allow a direct access to the Casimir force driven heat flux.  Although the experiment is impressive, it is only able to measure the Casimir force between two membranes at different bath temperatures. Actually, our detailed analysis makes clear that the temperatures of the membranes only changes by a vanishingly small amount and it might be very hard if not impossible to measure it, because it is just so extremely small ($0.02\times10^{-15}\,{\rm K}$) that it is not accessible with existing temperature measurement devices. Even if one could measure any temperature changes it would mainly be caused by the radiative heat flux ($\sim0.02\,{\rm K}$) which is 15 orders of magnitude larger and mask the contribution of the Casimir force as can be seen in Fig.~\ref{Fig:HeatFlux}. It is therefore clear, that the Casimir force driven heat flux does not provide any new route for thermal management at the nanoscale, but is rather of fundamental interest and so far unmeasured. Finally, the claim, that the effect results in large temperature changes of the membranes is only true for the mode temperatures which are no thermodynamical temperatures, i.e.\ they cannot be measured with any kind of thermometer. As discussed in detail, the full experiment can be understood without introducing the mode temperatures which helps to better understand what is really measured in the experiment, which is the change of mean square of the displacement of the membranes due to the Casimir force caused by the change in coupling with separation and not a real temperature change.

\acknowledgments
\noindent

S.-A.\ B.\ acknowledges support from Heisenberg Programme of the Deutsche Forschungsgemeinschaft (DFG, German Research Foundation) under the project No.\ 404073166. This project has received funding from the European Union's Horizon 2020 research and innovation programme under grant agreement No.\ 766853.


\begin{thebibliography}{99}
  \bibitem{FongEtAl} K. Y. Fong, H.-K. Li, R. Zhao, S. Yang, Y. Wang, and X. Zhang, Nature {\bf 576}, 243 (2019).
%  \bibitem{SpaceHeaters} https://www.scientificamerican.com/article/space-heater-scientists-find-new-way-to-transfer-energy-through-a-vacuum/ .
  \bibitem{PhysicsToday} Physics today, DOI:10.1063/PT.6.1.20191217a .
  \bibitem{Fiorino} A. Fiorino, D. Thompson, L. Zhu, B. Song, P. Reddy, and E. Meyhofer, Nano Lett. {\bf 18}, 3711-3715 (2018)
  \bibitem{SAB} S.-A. Biehs and G. S. Agarwal, J. Opt. Soc. Am. B {\bf 30}, 700 (2013).
  \bibitem{Pendry} J. B. Pendry, K. Sasihithlu, and R. V. Craster, Phys. Rev. B {\bf 94}, 075414 (2016).
  \bibitem{Joulain} Y. Ezzahri and K. Joulain, Phys. Rev. B {\bf 90}, 115433 (2014).
  \bibitem{Budaev} B. V. Budaev, D. B. Bogy, Appl. Phys. Lett. {\bf 99}, 053109 (2011).
  \bibitem{Reddy} L. Cui, S. Hur, Z. A. Akbar, J. C. Kl\"{o}ckner, W. Jeong, F. Pauly, S.-Y. Jang, P. Reddy, and E. Meyhofer, Nature {\bf 572}, 628 (2019).
   \bibitem{PvH} D. Polder and M. van Hove, Phys. Rev. B {\bf 4}, 3303 (1971).
\end{thebibliography}
\end{document}